\documentclass[aps,prb,twocolumn,cite,dcolumn,groupedaddress,reprint,nofootinbib,letterpaper]{revtex4-1}
\usepackage{amsmath}
\usepackage{graphicx}% http://ctan.org/pkg/graphicx
\usepackage{epsfig}
\usepackage{float}
\usepackage[textwidth=2.5cm]{todonotes}

\usepackage{rotating}
\usepackage{amsxtra}
\usepackage{url}
\usepackage{txfonts}
\usepackage{mathrsfs}
\usepackage{booktabs}
\usepackage{caption}
\usepackage{subfig} 
\usepackage{dcolumn}
\providecommand{\abs}[1]{\lvert#1\rvert}

\begin{document}

% Use the \preprint command to place your local institutional report
% number in the upper righthand corner of the title page in preprint mode.
% Multiple \preprint commands are allowed.
% Use the 'preprintnumbers' class option to override journal defaults
% to display numbers if necessary
%\preprint{}

%Title of paper
\title{The finite element method applied to the study of two-dimensional photonic crystals.}

% repeat the \author .. \affiliation  etc. as needed
% \email, \thanks, \homepage, \altaffiliation all apply to the current
% author. Explanatory text should go in the []'s, actual e-mail
% address or url should go in the {}'s for \email and \homepage.
% Please use the appropriate macro foreach each type of information

% \affiliation command applies to all authors since the last
% \affiliation command. The \affiliation command should follow the
% other information
% \affiliation can be followed by \email, \homepage, \thanks as well.

%\author{Imanol Andonegui and Angel J. Garcia--Adeva}
\author{Imanol Andonegui}
\email{imanol\_andonegui@ehu.es}
\author{Angel J.~Garcia-Adeva}
\email{angel.garcia-adeva@ehu.es}

%\homepage[]{Your web page}
%\thanks{}
%\altaffiliation{}
\affiliation{\\\\Departamento de Fisica Aplicada I, E.T.S. Ingenieria de Bilbao,\\ Universidad del Pais Vasco (UPV/EHU),\\ Alda.~Urquijo s/n, 48013 Bilbao, Spain}

%Collaboration name if desired (requires use of superscriptaddress
%option in \documentclass). \noaffiliation is required (may also be
%used with the \author command).
%\collaboration can be followed by \email, \homepage, \thanks as well.
%\collaboration{}
%\noaffiliation

\date{\today}

\begin{abstract}
Calculations of the photonic band structure, transmission coefficients, and quality factors of various two-dimensional, periodic and aperiodic, dielectric photonic crystals by using the finite element method (FEM) are reported. The fundamental equations governing the propagation of electromagnetic waves in inhomogeneous media are revisited together with the boundary conditions required for each of the performed calculations. A detailed account of the eigenvalue and harmonic propagation analysis of the electromagnetic problem is reported for several periodic and finite-length structures. It is found that this method reproduces quite well previous results for these lattices obtained with the standard plane wave method with regards to the eigenvalue analysis (photonic band structure calculations). However, in contrast with frequency methods, the finite element method easily allows one to study the time-harmonic propagation of electromagnetic fields and, thus, to calculate the transmission coefficient of finite clusters in a natural way. Moreover, the advantages of using this real space method for structures of arbitrary complexity are also discussed. In addition, point defect cluster quality factor calculations are reported by means of FEM and they are compared to the ones obtained with the FDTD and \texttt{Harminv} methods. As a result, FEM comes out as an effective, stable, robust, and rigorous tool to study light propagation and confinement in both periodic and aperiodic dielectric photonic crystals. 

\end{abstract}

% insert suggested PACS numbers in braces on next line
%\pacs{}
% insert suggested keywords - APS authors don't need to do this
%\keywords{}

%\maketitle must follow title, authors, abstract, \pacs, and \keywords
\maketitle

% body of paper here - Use proper section commands
% References should be done using the \cite, \ref, and \label commands
\section{Introduction}
Photonic crystals have generated a surge of interest in the last decades because they offer the possibility to control the propagation of light to an unprecedented level \cite{Joannopoulos1995,Sakoda2001,cefe,istrate2006}. In its simplest form, a photonic crystal is an engineered inhomogeneous periodic structure made of two or more materials with very different dielectric constants. When an electromagnetic wave (EM) propagates in such a structure whose period is comparable to the wavelength of the wave, unexpected behaviors occur. Among the most interesting ones are the possibility of forming a complete photonic band gap (CPBG) \citep{Angel1,Angel2}, that is, a frequency range for which no photons having frequencies within that range can propagate through the photonic crystal (PC), to localize light by introducing several types of defects in the lattice, or enhancing certain non-linear phenomena due to small or anomalous group velocity effects. Unfortunately, all these nice features come at a price: the length scales required in order to fabricate a photonic crystal appropriate for operation at telecommunication frequencies are below the micron, so that ingenious innovations were required in order to actually fabricate such structures. Furthermore, fabricating devices based on these lattices is even more challenging from a technological point of view, especially in three dimensions . This has resulted in a lot of effort being devoted to investigating photonic crystal devices based on two-dimensional heterostructures, such us three-dimensional waveguides made of a two-dimensional photonic crystal core sandwiched between two layers of substrate that confines light by simple refraction index matching. Therefore, investigating two-dimensional photonic crystals is not a mere academic exercise but an important task, both at the fundamental and applied levels.
From a theoretical perspective, one of the most important venues of research is to develop numerical methods to solve Maxwell's equations of an EM wave propagating in a PC that are reliable, fast, and capable of dealing with large systems as close as possible to the real ones employed for experiments. The reason why numerical methods are so important in this field is rooted on the fact that the predictions of Maxwell's equations for these systems are in excellent qualitative and quantitative agreement with experiments and they are able to describe the ample phenomenology exhibited by these systems. The development of these numerical methods has undergone an evolution that mimics to a certain extent the one underwent by the techniques used to calculate the electronic band structure of semiconductors. Thereby, while early research efforts focused on ideal infinite periodic photonic crystals, for which techniques on the frequency space are ideally suited (such as the plane wave method (PW) \cite{Joannopoulos1995,Yablo1991,Johnson2001} or algorithms based on the tight binding method), it was soon realized that the real interest of these systems is disordered photonic crystals. Frequency methods can be used to address the calculation of certain quantities (such as the photonic band structure or density of states) for some types of disorder, mainly point or line defects, by performing supercell calculations \cite{mpbman} but they are not very useful for studying disorder that is not localized (such as random displacements of the dielectric constituents of the photonic crystal or compositional disorder), which are of great practical interest, as these types of disorder are usually associated to the fabrication techniques themselves. Furthermore, frequency methods, in principle, are only applicable to infinite systems whereas real systems are always finite. For these reasons, people have started to pay attention during the last years to real space methods. These methods differ from the previous ones in that they work with a finite cluster which does not need to be periodic. 

 Interestingly, a method that has received very little attention in this field even though is has been known and extensively used in other areas of Physics and Engineering is the finite element (FE) method \cite{fem01,Lin2007,Frei2004,Sopaheluwakan2003,Pomar2004}. Indeed, there are a number of reasons that seem to suggest that it could be applied well to the study of the propagation of EM waves in inhomogeneous media. In particular, this method allows one to study geometries of arbitrary complexity, it can deal with frequency dependent dielectric functions (metallic inhomogeneous structures) in a natural way, discontinuities in the dielectric function are not especially detrimental for convergence of the method, and the quantities are already calculated in the stationary regime. The only shortcoming of the method is its extensive computer memory usage. However, the demand of this resource is quite insensitive to the presence of defects, which could render this technique very advantageous for studying disordered lattices \cite{Sopaheluwakan2003}. Therefore, at the present status of the photonic band gap materials field, it is important to assess whether this method could be useful to investigate the properties of photonic crystals from both the fundamental and applied points of view. This is precisely one of the aims of this paper: we report a comprehensive study of both the band structure and transmission coefficients of various two-dimensional photonic crystals using the finite element method.  The other main goal is to study the feasibility of this method to analyze resonant cavities in photonic crystals and their associated quality factors.

The structure of the paper is as follows: in the next section we revisit the equations that describe the propagation of electromagnetic fields in two-dimensional inhomogeneous media and, in particular, in photonic crystals. Also, the boundary conditions required for each particular simulation are presented. Section \ref{square-tri-lattice} summarizes the results for topologies with the symmetry of the square and triangular lattices. Section \ref{sec_periodicdefect} examines the dispersion relations for crystals with point defects for both the square and triangular lattices. Then, defect mode tunability is briefly discussed. Section \ref{sec_finitedefect} addresses finite extent point defect PCs wherein issues such as the effect of supercell dimensionality in the formation of photonic localized states and, hence, the improvement of its quality factor are discussed. Finally, in section \ref{conclusions}, we state the conclusions of this work.

\section{The mathematics behind the propagation of EM waves in inhomogeneous media}\label{sec_theory}

The propagation of an electromagnetic wave through an inhomogeneous medium (such as the photonic crystals considered in this chapter) is described by Maxwell's equations. In particular, for two-dimensional media, any electric or magnetic field can always be expressed as a linear combination of a transversal electric (TE) and a transversal magnetic fields (TM). In the first case, the electric field is perpendicular to the photonic crystal plane --whereas the magnetic field is constrained into this plane-- and the sourceless Maxwell's equations are reduced to a Hemholtz equation for the electric field given by
\begin{equation}\label{tewe}
 	\nabla^2 E_z (\vec{r}) + k_0^2 \epsilon_r(\vec{r}) E_z (\vec{r})=0,
\end{equation}
where $E_z(\vec{r})$ is the $z$-th component of the electric field at position $\vec{r}$, $\epsilon_r(\vec{r})$ is the inhomogeneous relative dielectric constant of the photonic crystal, and $k_0=\omega/c$ with $\omega$ the angular frequency of the incident electric field and $c$ the speed of light in free space. In writing down equation \eqref{tewe}, it has been assumed that the photonic crystal is non-magnetic ($\mu_r = 1$) and non-conducting ($\sigma=0$). Once equation \eqref{tewe} is solved, the time-harmonic electric and magnetic fields are easily calculated as
\begin{align}
 	\vec{E}(\vec{r},t)&=E_z (\vec{r})\,e^{-i\omega t}\,\hat{z}\\
	\vec{H}(\vec{r},t)&=\frac{i}{k_0}\nabla\times\epsilon_r(\vec{r})\vec{E}(\vec{r},t).
\end{align}
On the other hand, in the case of TM polarization, the magnetic field is perpendicular to the photonic crystal plane --whereas the electric field is constrained into this plane-- and the sourceless Maxwell's equations reduce to a Hemholz equation for the magnetic field given by
\begin{equation}\label{tmwe}
	\nabla\left(\frac{\nabla H_z (\vec{r})}{\epsilon_r(\vec{r})}\right)-k_0^2 H_z (\vec{r})=0,
\end{equation}
where $H_z (\vec{r})$ is the $z$-th component of the magnetic field at position $\vec{r}$. The time-harmonic electric and magnetic fields are easily calculated once equation \eqref{tmwe} is solved and they are given by
\begin{align}
	\vec{H}(\vec{r},t) &= H_z (\vec{r})\,e^{-i\omega t}\,\hat{z}\\
	\vec{E}(\vec{r},t) &= -\frac{i}{k_0 \epsilon(\vec{r})} \nabla\times\vec{H}(\vec{r},t).
\end{align}
An important aspect of any electromagnetic simulation is the use of appropriate boundary conditions at the interfaces. On the one hand, for the photonic band structure calculations reported below, it is necessary to implement boundary conditions that mimic an infinite simulation domain together with the periodicity of the photonic crystal lattice. An infinite medium is simulated by using perfect magnetic (PMC) or perfect electric conductor (PEC) boundary conditions that mirror the simulation domain. The former condition is used for TE polarization of the EM field and ensures that the component of the magnetic field tangent to the boundary is identically zero at the outer interface, that is,
\begin{equation}
 \hat{n}\times\vec{H}=\vec{0},
\end{equation}
where $\hat{n}$ is a unit vector perpendicular to the outer simulation domain surface at each point. The later is used for TM polarization of the EM field and ensures that the component of the electric field tangent to the boundary is identically zero at the outer interface, that is,
\begin{equation}
 \hat{n}\times\vec{E}=\vec{0}.
\end{equation}
The periodicity of the photonic crystal lattice is ensured by an adequate use of Bloch's theorem at the boundaries of the photonic crystal unit cell. This theorem states that when the electric (magnetic) field propagates from one point in the PC to another one separated from the previous one by a lattice vector, $\vec{R}$, the only effect on the EM field is a change of its phase,
\begin{equation}
	E_z (\vec{r}+\vec{R}) = e^{i\,\vec{k}\cdot\vec{R}} E_z (\vec{r})\label{eq_bloch_ez}
\end{equation}
and
\begin{equation}
	H_z (\vec{r}+\vec{R}) = e^{i\,\vec{k}\cdot\vec{R}} H_z (\vec{r})\label{eq_bloch_hz}
\end{equation}
for TE and TM polarization, respectively. In these expressions, $\vec{R}$ is a vector of the photonic crystal lattice and $\vec{k}$ is the wavevector of the electromagnetic wave. On the other hand, for the transmittance calculations reported below, finite clusters in the direction parallel to the incident wave vector were used, whereas the clusters are of infinite extension in the perpendicular direction. In order to mimic such a material, PMC and PEC boundary conditions were used for the outer interfaces that limit the simulation domain in the direction perpendicular to the incident wave vector for TE and TM polarization, respectively. For the interfaces at which the EM wave enters and exits the cluster, the situation is a little more involved because it is necessary to avoid unphysical reflections due to the finite size of the cluster and, thus, perfectly matched layers were used at these interfaces. The equations that describe such boundaries are given by
\begin{align}
\hat{z}\cdot\hat{n}\times(\nabla\times E_z\hat{z})-i\,\beta E_z &=-2 i \beta E_{0z}\\
\hat{z}\cdot\hat{n}\times(\nabla\times H_z\hat{z})-i\,\beta H_z &=-2 i \beta H_{0z},
\end{align}
where $E_{0z}$ and $H_{0z}$ are the initial values of the electric and magnetic fields at the boundaries, respectively, and $\beta=k_0$ is the propagation constant. The upper condition applies to TE polarization, whereas the lower one corresponds to TM polarization. If the electric field is an eigenmode of the boundary, the boundary is exactly non-reflecting.

\section{Simulation results for photonic crystals based on square and triangular lattices}
\label{square-tri-lattice} 

\begin{figure*}
   \includegraphics{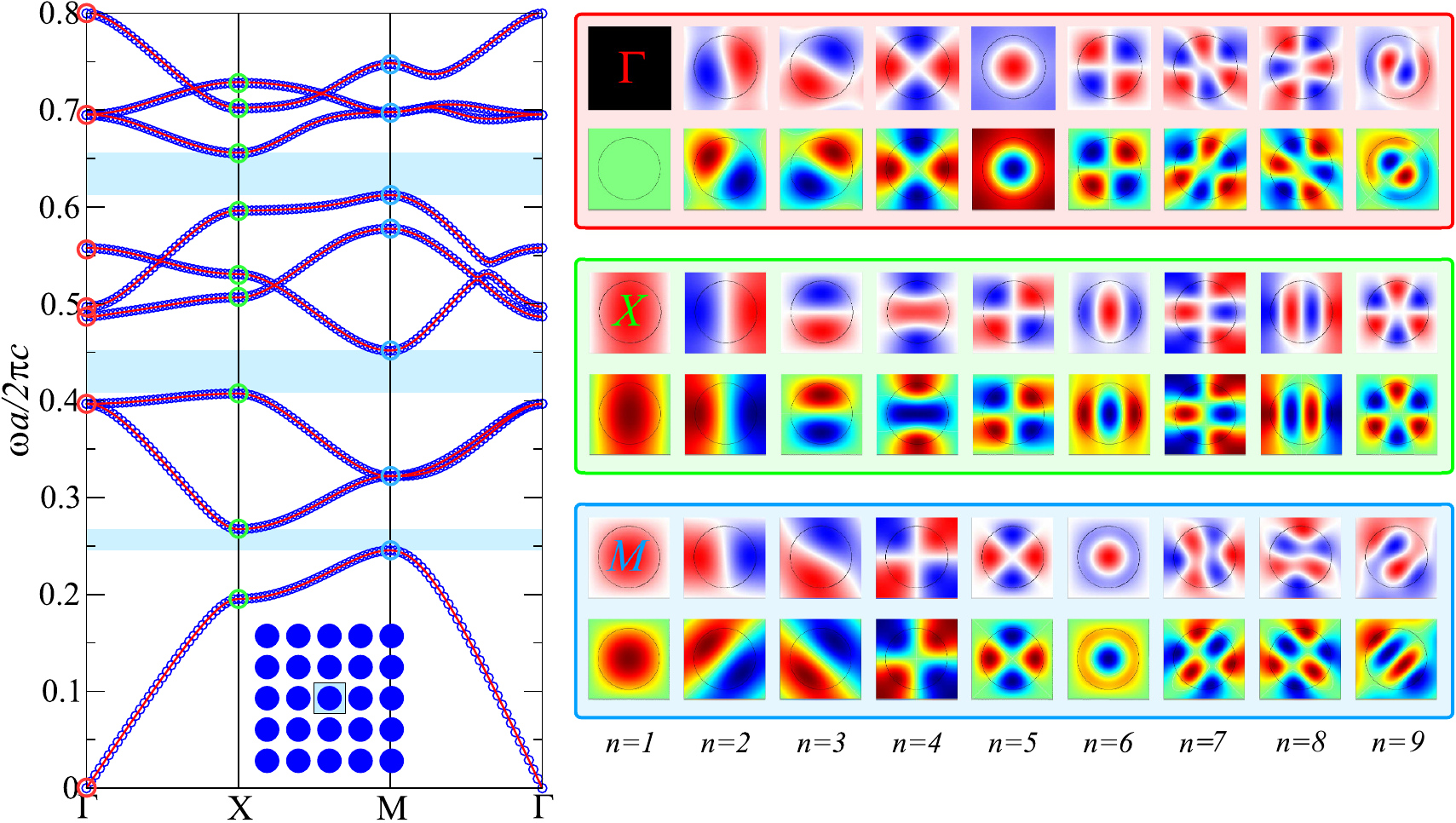}
	\caption{\label{fig_perfect_square_lattice_bands}(color online). On the left: band structure calculated with the MPB (blue circles) and COMSOL (red line) software packages for TE-polarized EM waves. On the right: $E_z$ patterns in the unit cell calculated with MPB (upper row in each rectangle) and COMSOL at the special symmetry points $\Gamma$, $X$, and $M$.}
%\end{wrapfigure}
\end{figure*}

The first structure analyzed in this work is a photonic crystal made of dielectric circles whose centers occupy the positions of a square lattice and is depicted in the inset of Fig.~\ref{fig_perfect_square_lattice_bands}. The dielectric material was assumed to be linear, isotropic, and non-magnetic \footnote{This applies to the rest of photonic crystals considered in the present work}. The dielectric constant of the rods has a value of $9$. The ratio $\frac{r}{a}$, where $r$ is the radius of the cylinders and $a$ the lattice parameter, was taken as $0.38$. This well known structure was first investigated by McCall and coworkers in order to compare the predictions of theory with experimental results with regards to the localization of light in strongly scattering media. In the present context, we have studied this topology in order to check how well the FE method fares when compared with the plane wave method that, as stated above, is the one commonly used to calculate photonic band structures. The first nine photonic bands were calculated for transversal electric (TE) polarization along the path that delimits the irreducible part of the 1st Brillouin Zone (1BZ). For the FE calculation, the square unit cell was divided in $3720$ mesh elements and periodic boundary conditions given by Bloch's theorem were implemented. For the MPB calculation, a resolution of $64\times64$ ($=4096$) grid elements were used and the dielectric constant was average over $9$ grid points. The resulting  photonic band structure is depicted in Fig.~\ref{fig_perfect_square_lattice_bands}. As it is customary, the dimensionless quantity $\omega \textit{a}/2\pi c = \textit{a}/\lambda$ has been used to characterize the frequency of the incident EM wave, where $\omega$ is the frequency of the incident EM wave and $\lambda$ the associated wavelength. The corresponding eigenmodes of the $z$-component of the electric field, $E_z$, were also calculated at the $\Gamma$, $M$, and $K$ points of the 1BZ and they are shown also in figure \ref{fig_perfect_square_lattice_bands}. It is clear from that figure that the band structure calculated with the FE method faithfully reproduces the one calculated with MPB to its minimum details. There are three photonic gaps in the band structure of this lattice whose sizes coincide with the calculated ones with MPB. Also, the modes calculated with the FE method closely resemble those calculated with MPB up to a trivial symmetry operation or linear combination of degenerate modes. In addition, a quantity that can be readily calculated with the FEM method is the transmittance of a finite photonic crystal. The transmittance of the cluster is calculated using the usual approach of power integration along a domain boundary. The boundary conditions for the simulation domain were set as follows: on the input and output boundaries (left and right boundaries) a perfectly matched layer was used to avoid spurious reflections from non-physical boundaries. The $z$-component of the electric field was set to $1$  and $0$ at the initial time of the simulation on the input and output boundaries, respectively. On the other hand, perfectly magnetic conductor boundary conditions that set the tangential magnetic component of the magnetic field to zero were used for the upper and lower boundaries \footnote{Simulations using different boundary conditions for the upper and lower boundaries were performed and no significant differences in the calculated transmission spectra were found} in order to mimic an infinite stripe in that direction.

\begin{figure*}
	
	\includegraphics{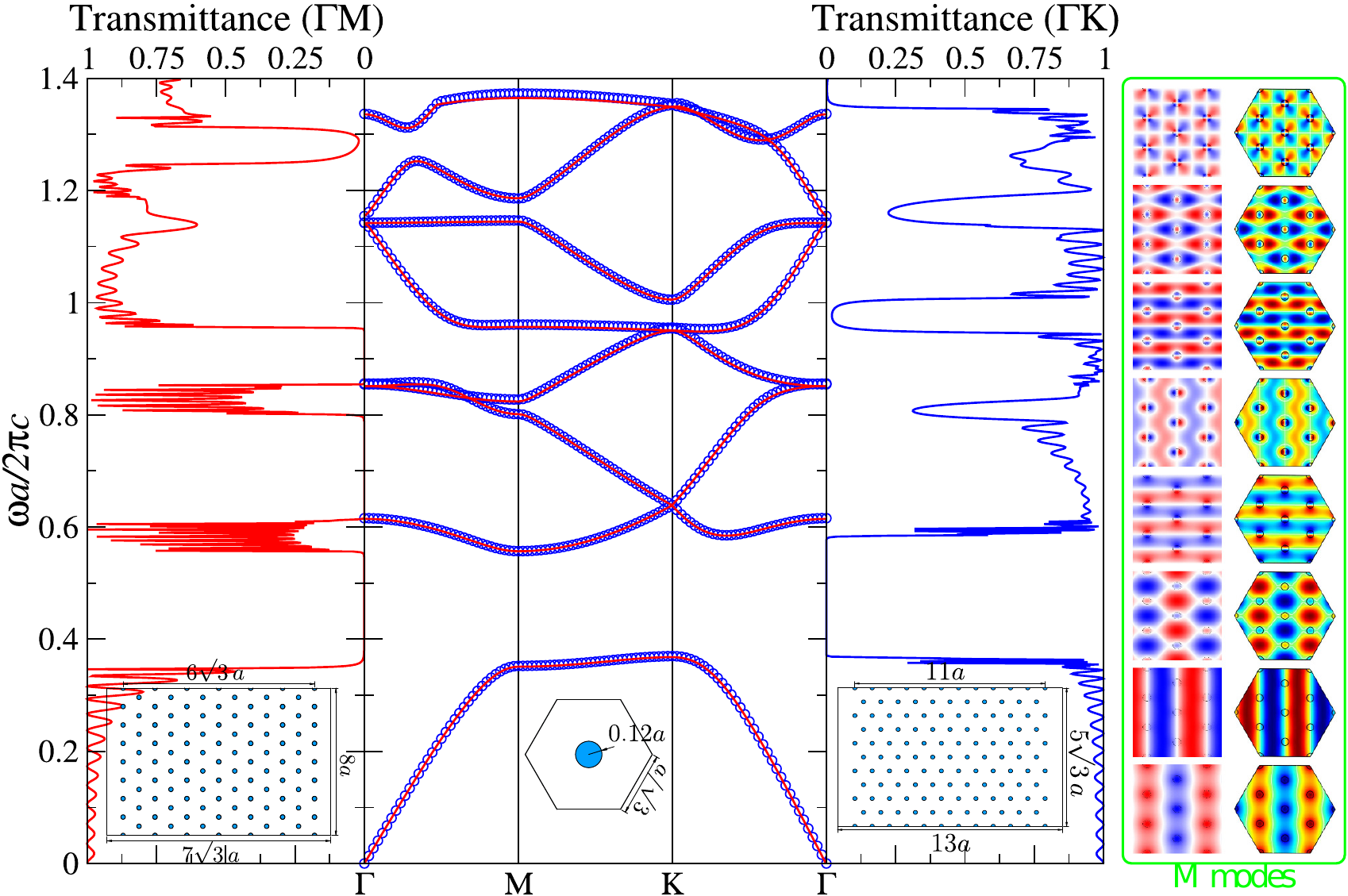}
	\caption{\label{fig_tdrtepol}(color online) Left panel: Transmittance for TE-polarized waves propagating in the $\Gamma M$ direction of the hexagonal photonic crystal cluster depicted in the inset. Central panel: Comparison between the band structure calculated with COMSOL (red line) and MPB (blue circles) for TE-polarized EM waves along the boundary of the irreducible part of the 1BZ. The inset shows the unit cell used for the calculations. Right panel: Transmittance for TE-polarized waves propagating in the $\Gamma K$ direction of the hexagonal photonic crystal cluster depicted in the inset. Inside the green rectangle: Comparison between the $E_z$ field patterns at the $M$ point calculated with MPB (on the left) and FEM (on the right) for the first eight bands (the band index increases from bottom to top). A portion of the photonic crystal that contains $3\times3$ unit cells is displayed in order to show the hexagonal symmetry of the modes.}
\end{figure*}

Another important class of two-dimensional photonic crystals is the one formed by those structures based on the triangular lattice because this is the symmetry most commonly used for practical applications \cite{Joannopoulos1995}. For this reason it is important to test whether the FE method produces correct results for this non-orthogonal lattice. A photonic crystal made of dielectric cylinders whose centers occupy the sites of an hexagonal lattice of lattice constant $\textit{a}$ (see the insets in Fig.~\ref{fig_tdrtepol}) was analyzed. A value of $12$ was used for the dielectric constant of the non-empty part of the photonic crystal. The radius of the cylinders is $0.12a$. The simulation setup was very similar to the one used for the square lattice: For the FE band calculation, the unit cell was divided into $1086$ mesh elements and periodic boundary conditions were implemented across the boundaries. For the transmittance calculation in the $\Gamma M$ direction, a cluster comprising $12\times8$ unit cells (dielectric rods) was divided into $60748$ mesh elements. For the transmittance calculation in the $\Gamma K$ direction, a cluster comprising $11\times10$ unit cells was divided into $67282$ mesh elements. The same boundary conditions as for the square lattice calculations described above were used. The photonic band structure, transmittances in the $\Gamma M$ and $\Gamma K$ directions, and $E_z$ field patterns for TE-polarized EM waves are reported in Fig.~\ref{fig_tdrtepol}. The band structure calculated with the FE method faithfully reproduces the one obtained with MPB. This structure possesses one gap between the first and second bands for TE-polarized EM waves that can be clearly seen in the transmission spectra for light propagating in both the $\Gamma M$ and $\Gamma K$ directions as wide opaque regions. There are also some partial gaps, as the ones occurring between the second and third bands and the fourth and fifth bands in the $\Gamma M$ direction that, however, are not present in the $\Gamma K$ direction. Also, there are some opaque regions not related to gaps on the band structure but rather to the existence of uncoupled modes, such as the one associated to the 5th band, that is uncoupled in the $\Gamma K$ direction.

\begin{figure}
		\includegraphics{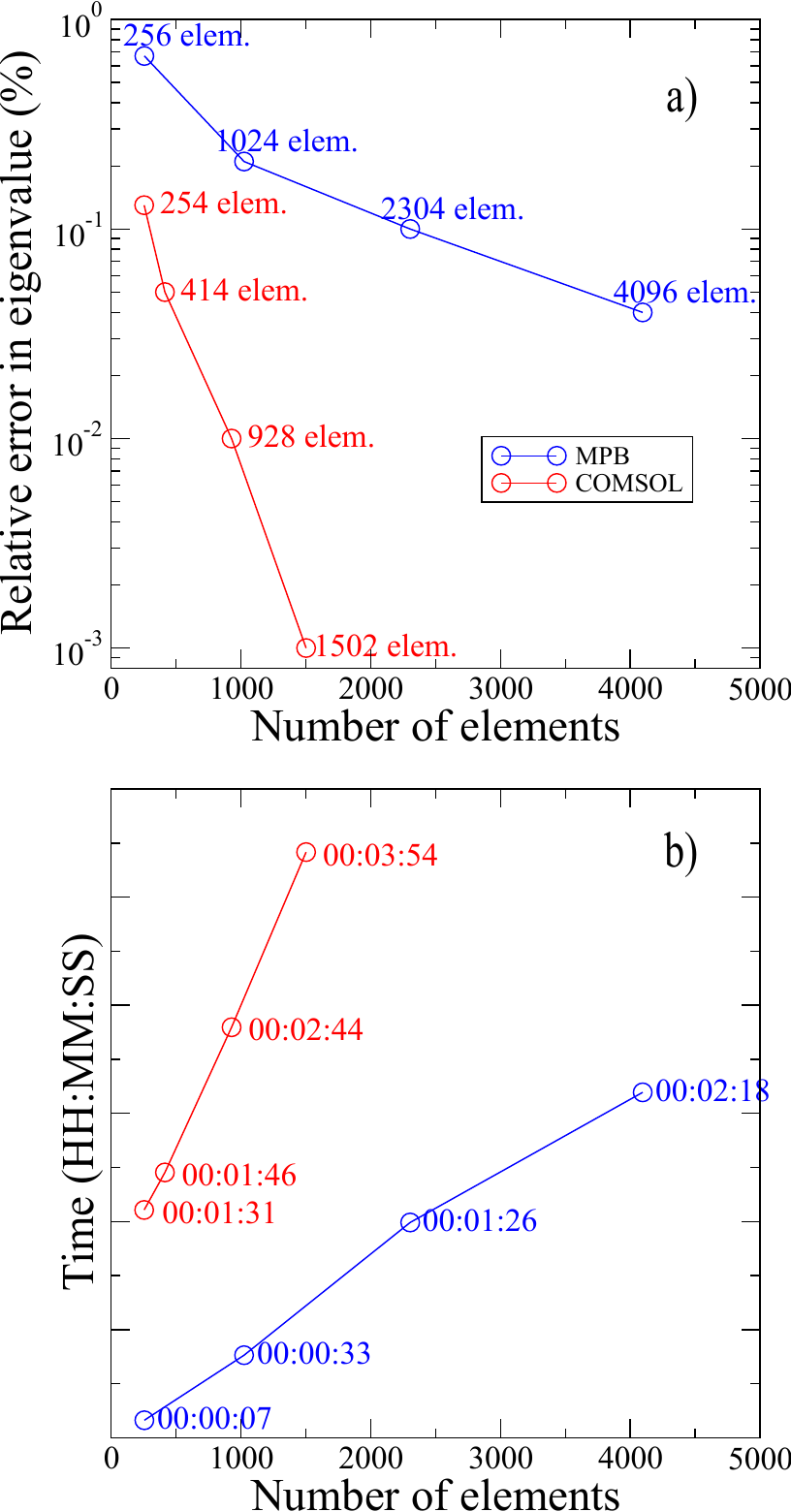}
	\caption{\label{fig_comparison_fem_mpb_square_lattice}(color online) (a) Evolution of the relative error with the number of elements in which the lattice is discretized. (b) Evolution of the simulation run time with the number of elements in which the lattice is discretized.}
\end{figure}

Even though that is not the main purpose of this paper, it is illustrative to compare to some extent the accuracy and speed of the FE calculations with the ones performed by using the PW method. To estimate the relative accuracy, we computed the percent error in the eigenvalue calculated at the $X$ point for the ninth band for different discretizations of the square unit cell. In the FE case, meshes with $254$, $414$, $928$, and $1502$ elements were used, whereas for the calculations done with MPB, resolutions of $16\times16$ ($=256$) grid elements, $32\times32$ ($=1024$) grid elements, $48\times48$ ($=2304$) grid elements, and $64\times64$ ($=4096$) grid elements were used. We took the eigenvalue calculated with MPB by using a resolution of $256\times256$ and a mesh size of $25$ as the exact one. The result of this comparison can be seen in Fig.~\ref{fig_comparison_fem_mpb_square_lattice}a. It is noticeable that the FE method gets a better accuracy with coarser discretizations of the lattice than the PW does. This is due to the use of second order Lagrange elements. However, the differences between both methods should be negligible for most applications. Of course, this better accuracy comes at a price and the simulation times for the FE calculations are longer by a noticeable factor than those for the MPB ones, as can be seen in Fig.~\ref{fig_comparison_fem_mpb_square_lattice}b, where we have depicted the evolution of the simulation runtime with the number of mesh elements and grid elements for the FE and MPB calculations, respectively.

\section{Defect states in 2D photonic crystal superlattices}\label{sec_periodicdefect}
In complete analogy with semiconductors, the physics of disordered photonic crystals is even richer and their potential applications even more promising. Among these, photonic crystals in which defects are placed in a controlled way forming point or line defects are especially important for telecommunication applications. Their importance is rooted on the fact that these types of defects can lead to the formation of localized states inside the photonic gap that allows one to localize light around the defects. 

Point defect photonic crystal configurations have been widely studied as promising candidates for the enhancement of strong coupling between the resonance cavity and quantum dots light emitters \cite{Edo} or to reduce the lasing threshold of a certain laser emitter \cite{Sibilia}. Besides, they have been described as key elements for several applications in many areas of physics and engineering such as enhancing high directivity antennas \cite{Temelkuran}, designing low power consumption and highly tunable optical buffering devices \cite{Tanabe, Baba}, or constructing new PC-VCSEL lasers \cite{Song} as well as for biosensing applications \cite{Rajesh}, just to cite some examples.

In the simulations reported in this section, periodicity of the lattice has been intentionally broken and hence a defect is introduced into the otherwise perfectly ordered dielectric distribution, giving rise to localized states within the band gap region. When a single rod is removed from the dielectric lattice, light bounces back and forth in the disordered area, trapped by the surrounding band gap, whilst no other leakage mechanism is present. The resulting photonic crystal can still be classified on the basis of unit cell calculations, as long as the fundamental periodic cell hosts sufficient rods around the imperfection to make sure that the defect state does not overlap with neighboring replicas of itself. Furthermore, in these single rod perturbed cells, rotational symmetry remains invariant and thus comparative studies with previous perfect square and triangular lattice based calculations can still be performed at the edge of the 1BZ.

\subsection{A square lattice of dielectric rods with a point defect: the McCall's experiment}\label{defect_sq}

\begin{figure*}
	\includegraphics{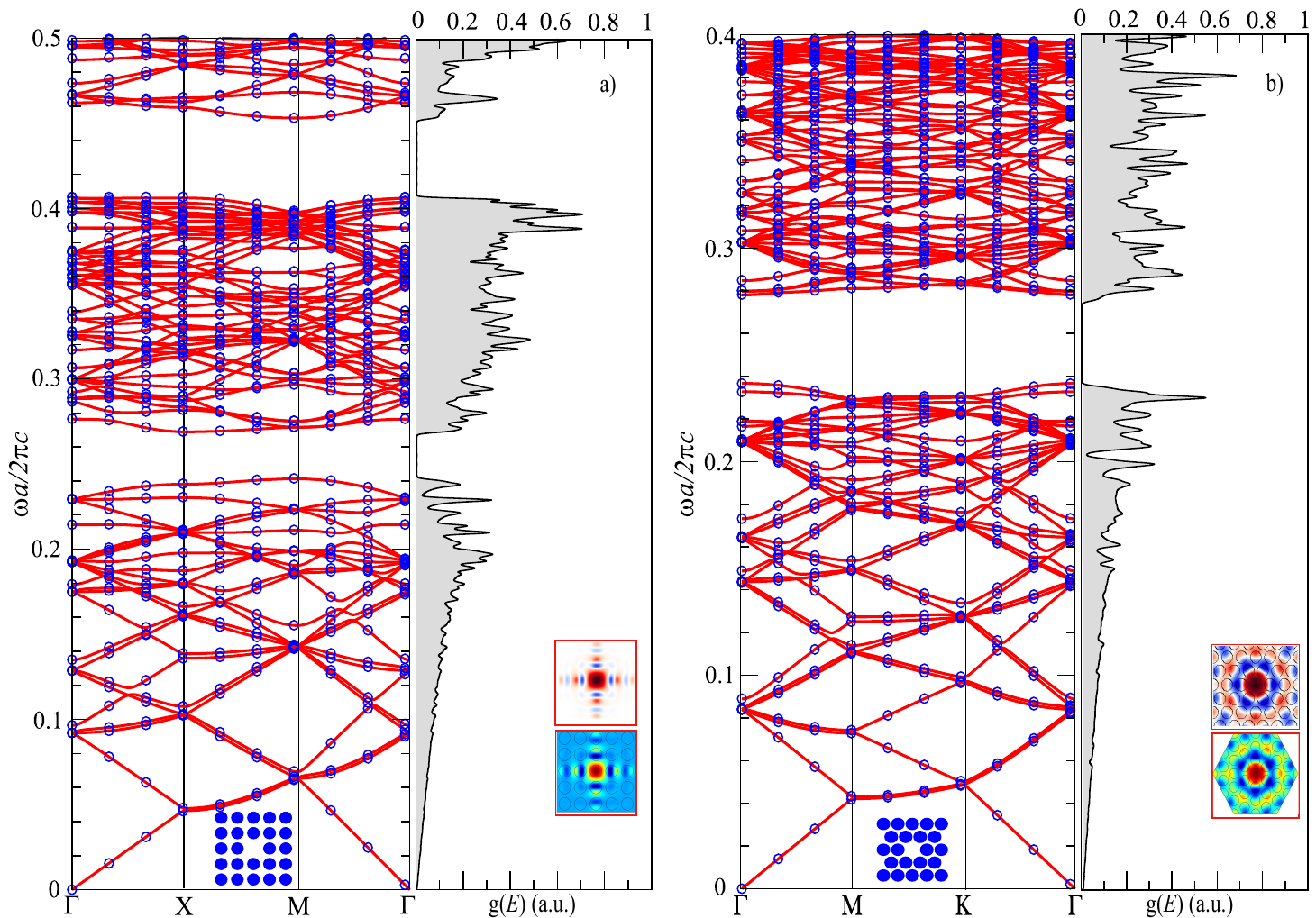}
	\caption{\label{finalsqtri038periodic2}(color online) a) On the left: Band structure of a $5 \times 5$ square lattice periodic supercell calculated with MPB (blue circles) and FEM (red lines), wherein a defect state has been excited around $\frac{\omega a}{2 \pi c}=0.466\pm 0.002$. On the right: density of photonic states (DOS), calculated by FEM. The band gap region is clearly distinguishable and a weakly localized defect mode merging from the upper band can be seen. In the inset, $E_{z}$ patterns for a $5 \times 5$ supercell calculated by both methods are shown for the M symmetry point.
b) On the left: band structure of a $5 \times 5$ triangular lattice calculated with MPB (blue circles) and FEM (red line). A single defect state merges at $\frac{\omega a}{2 \pi c}=0.285\pm 0.003$. On the right: density of photonic states, calculated by FEM. In the inset, $E_{z}$ patterns for the $\Gamma$ symmetry point. An hexagonal supercell has been used for the FEM calculation .}
\end{figure*}

The experimental evidence of spatial mode localization in an ordered dielectric lattice of rods in an air background was first given by McCall et al. \cite{McCall}. In the present context, we have studied this topology by FE method and compared it to the predictions of the PWE method \cite{Johnson2001}. Figure \ref{finalsqtri038periodic2} depicts the resulting dispersion diagram calculated for the experimental setup raised by McCall, where a single rod has been suppressed amidst the square lattice of dielectric rods, for which the normalized rod radius is $0.38\textit{a}$. The rod dielectric constant has been set to $9$, as in the original McCall's work. A $5 \times 5$ supercell has been used for the FEM calculations reported in this section. It has been discretized into $95903$ mesh elements. A $3096$ element grid has been used for the corresponding MPB calculations. The real part of the z-component of the electric field for the defect mode located at $a/\lambda=0.4686$ was calculated at the $\Gamma$ point of the 1BZ.
The main drawback of the supercell approach is the band-folding effect: redundant bands of the unit cell are folded back $N$ times ($N$ being the linear dimension of the supercell). This fact leads to larger computational times since the amount of eigenvalues to be solved grows $\sim N$. We have therefore solved eigenvalues for $211$ points in $K$-space for $100$ bands by both methods. Experimental results obtained by McCall and coworkers, MPB calculations, and FEM results fully agree as it is shown in Figs.~\ref{finalsqtri038periodic2} and \ref{finalsqtri017periodic2}.

Two modes make their way across the upper band gap region but only the defect mode depicted in Fig.~\ref{finalsqtri038periodic2}a concentrates its energy around the missing rod.
 
The density of states (DOS) shows that perturbation of a single rod induces a series of sharp Dirac peaks centered at the frequencies where the defect state occurs. DOS calculations have been also computed via FEM by randomly sampling k-points constrained to the irreducible portion of the IBZ for each lattice. In the long wavelength limit, this quantity clearly exhibits the linear behavior expected for propagation in an homogeneous two-dimensional dielectric medium.

\subsection{A triangular lattice of dielectric rods with a point defect}\label{defect_tri}
An analogous situation can be achieved by removing a single rod in a photonic crystal based on the triangular lattice. 

Fig.~\ref{finalsqtri038periodic2}b shows the corresponding results for the photonic crystal parameters used by Smith et al. ~\cite{Smith} in their investigation of the defect mode structures in the square and triangular lattices. Rod radii and dielectric constant parameters of the triangular lattice supercell are the same as in the orthogonal case of Fig.~\ref{finalsqtri038periodic2}a and, once again, FEM calculation and results obtained by means of PWE method coincide. The simulation setup was similar to the previously described one for the square lattice, but this time discretization of the supercell has been set to $71376$ mesh elements. Furthermore an hexagonal supercell has been used for the FEM calculations, as its shape matches the reciprocal lattice of a periodical triangular arrangement. 
In both cases, square and triangular lattice point defect supercells, the electric field $z$-component is well localized around the defect neighborhood and rapidly decays to small amplitudes as one moves further from the defect. Also, the eigenfunction around the lattice irregularity clearly shows an inherent symmetry. Indeed, both lattices present a monopole pattern with a single nodal plane through each dielectric rod \cite{Joannopoulos1995}. The symmetry of such point defect modes is analyzed in detail in \cite{Kim}.

\begin{figure*}
	\includegraphics{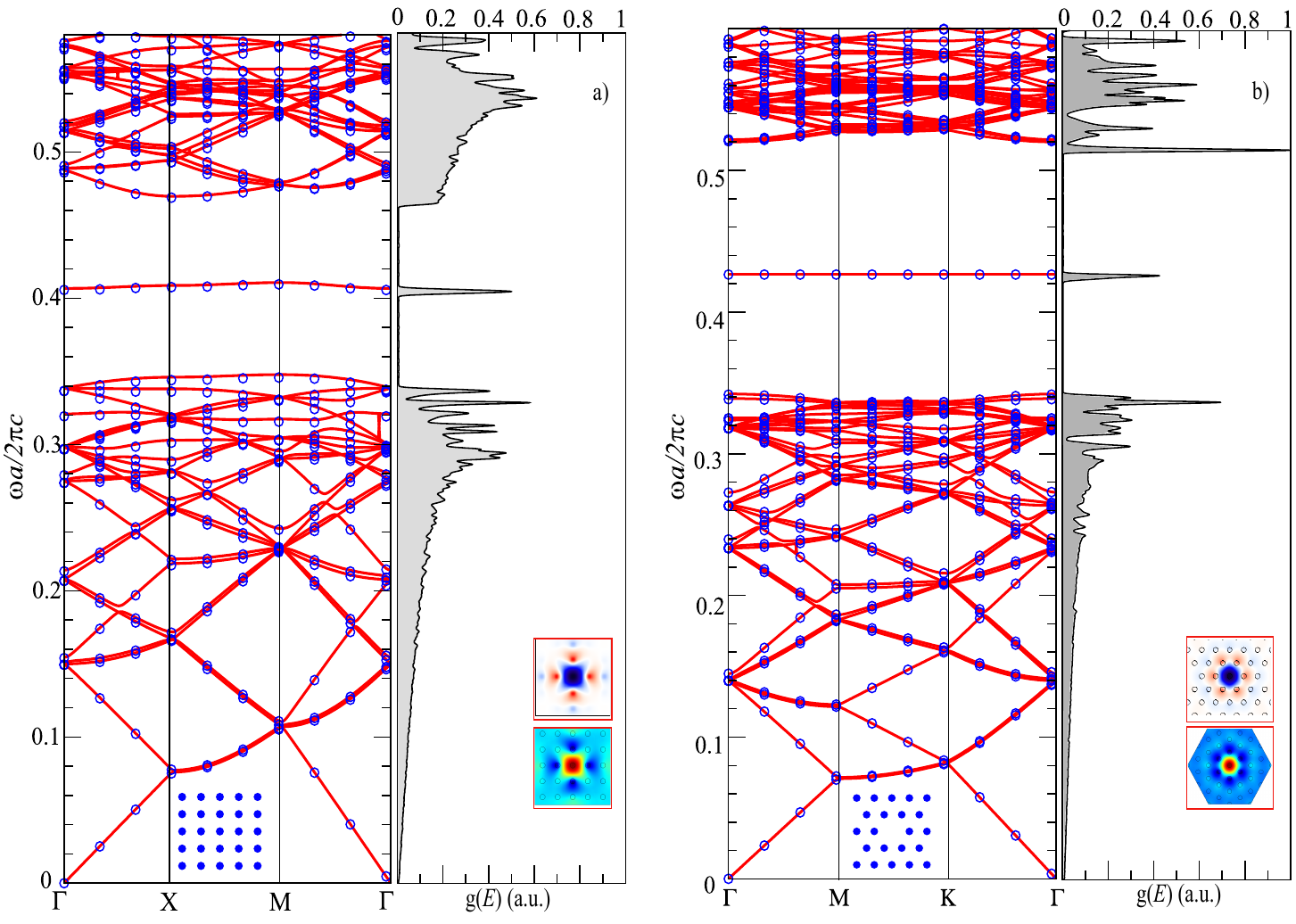}
	\caption{\label{finalsqtri017periodic2}(color online) a) On the left: band structure of a $5 \times 5$ square lattice periodic supercell calculated with MPB (blue circles) and FEM (red lines). The rod radii has been set to $0.17\textit{a}$, which ensures a better localization of the unique defect mode. On the right: density of photonic states, calculated by FEM. The defect mode clearly shows up in the gap region, since the defect state is strongly confined. In the inset, the $E_{z}$ patterns for a $5\times 5$ supercell calculated with both methods are shown for the M symmetry point. All these patterns match the ones presented in Fig.~\ref{finalsqtri038periodic2}a but the modal volume decreases significantly for $0.17\textit{a}$ rod lattices.
b) Triangular lattice made of $0.17\textit{a}$ rods where the central rod has been removed. The band gap increases and so it does the defect mode localization around the point defect.}
\end{figure*}

The confinement of the electric field amplitude in the cavity is directly related to the band gap strength. Stronger mode localization has been observed for wider gap geometries. This is a highly desirable effect since the gap-mid-gap ratio can be easily tuned by adequately preselecting rod radii and dielectric contrast. For rods of dielectric contrast equal to $9$ in square and triangular lattices, one can maximize the gap-mid-gap ratio for TE polarization by just adjusting the rod radii to $0.17\textit{a}$. Band diagrams and DOS calculations for $0.17\textit{a}$ radius rods for square and triangular lattice are reported in Fig.~\ref{finalsqtri017periodic2}a and Fig.~\ref{finalsqtri017periodic2}b. There, localized defect bands are strongly confined by a wider band gap. Moreover, these defect modes can be spatially translated into frequency by simply choosing the appropriate index contrast between lattice rods and point defect rod \cite{Joannopoulos1995}. With regards to this mode tunability some important properties of photonic crystal cavities can be conveniently enhanced. Indeed, incremental design procedures have successfully been demonstrated by means of algorithmic techniques rather than using intuitive trial and error design methods, where the defect dielectric function is tuned according to design requirements \cite{Cox,Shen,Geremia}. These facts will be discussed elsewhere \cite{unpublished_im}.

\section{Defect states in finite 2d photonic crystal clusters}\label{sec_finitedefect}

All the structures studied so far had an infinite extension in all space directions. Nevertheless, an engineered photonic crystal must be finite sized. This fact limits defect mode confinement factor as stored energy will be finally radiated outside the cavity. This circumstance gives rise to a quantity of very much practical importance: the quality factor of a photonic crystal resonator, \textit{Q}, which is just the ratio between the stored energy in the photonic crystal and the radiated energy per cycle. In the present context it is unnecessary to consider the radiated energy in the off-axis direction because the crystal has still infinite length in this direction, but inasmuch as the in-plane propagation will be limited to \textit{N} periods of dielectric rods, electromagnetic fields will still be leaked outwards. Quantitatively, these energy decay mechanisms are uncorrelated to each other, and so they can be studied separately. On the other hand, in the present calculations no losses, absorptions, nor imperfections have been taken into account and, consequently, the quality factor is only limited by the inherent energy leakage from the cavity to the radiative modes.

\begin{figure*}
	\centering
	\includegraphics[width=\textwidth]{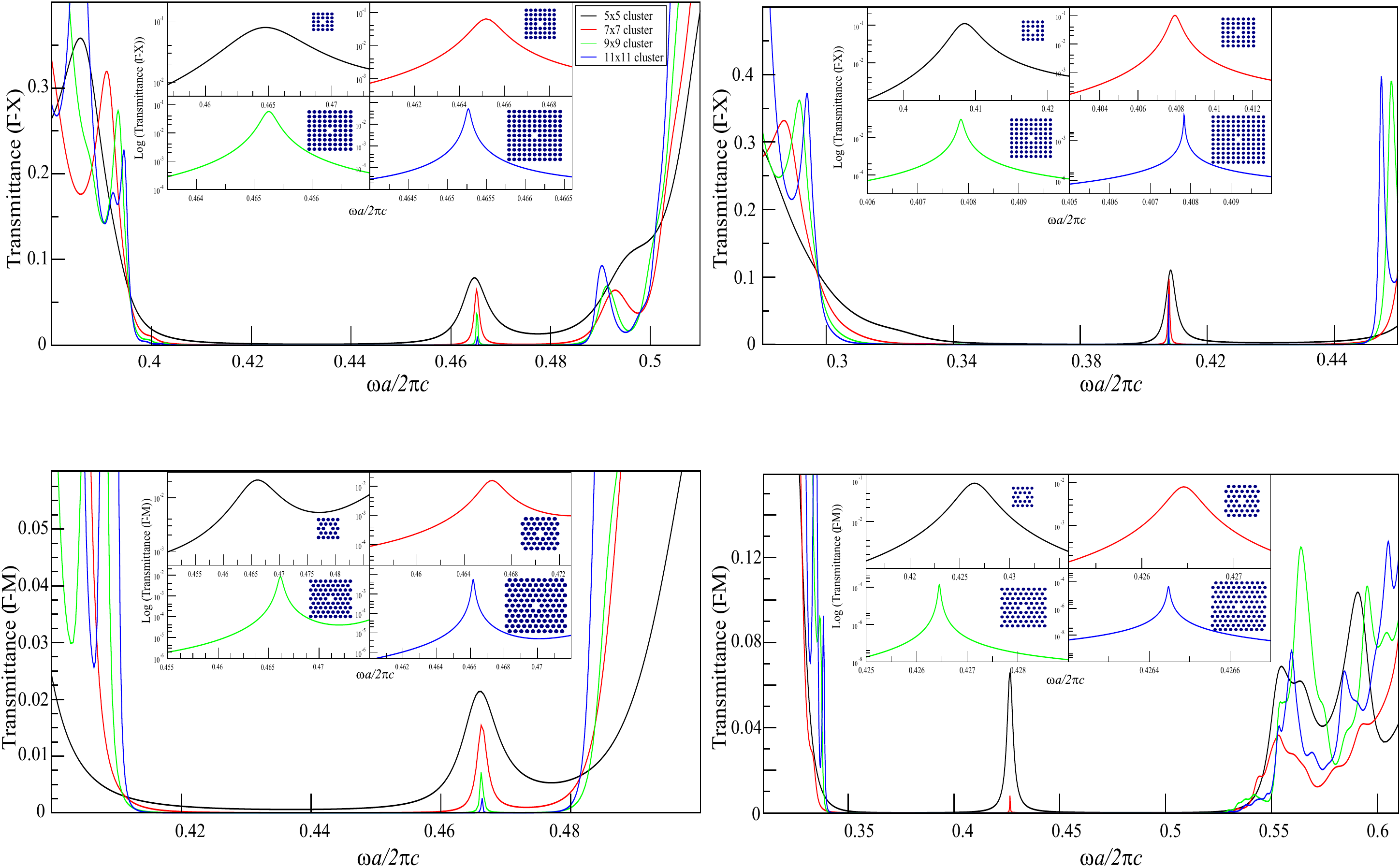}
    \caption{\label{transmittancesqtri}(color online) a) Top: transmittance diagrams for different cluster dimensions in a rectangular arrangement of dielectric rods with a central defect. On the left, rod radii is set to $0.38\textit{a}$ while on the right $\frac{r}{\textit{a}}$ is $0.17\textit{a}$. In the inset, detailed defect mode curves are depicted.(b) Down: analogous transmittance diagrams in a triangular cluster.}
\end{figure*}

Transmittance diagrams have been obtained using the FEM but, in contrast with the transmittances reported above for ideal lattices, in this section a finite length design is regarded. Therefore, PMC and PEC boundary conditions were substituted by PML interfaces in the direction perpendicular to the incident wave vector as well as at the interfaces at which the EM wave enters and exits the structure. Accordingly, transparent influx conditions were imposed to the outer boundaries of the finite cluster and a TE polarized source propagating in the $\Gamma-X$ direction was considered. Spurious reflections are thus prevented from being injected back into the simulation domain by locating these artificial perfectly matched layers all around the simulation domain boundaries. The equations that describe such boundaries are given by
\begin{align}
	\vec{z}\cdot\hat{n}\times(\nabla\times E_z\hat{z})-i\,k_0 E_z &=-i k_0 (1-\vec{k}_0 \cdot\hat{n}) E_{0z} e^{-i\vec{k}_0 \cdot \vec{r}}\\
	\vec{z}\cdot\hat{n}\times(\nabla\times H_z\hat{z})-i\,k_0 H_z &=-i k_0 (1-\vec{k}_0 \cdot\hat{n}) H_{0z} e^{-i\vec{k}_0 \cdot \vec{r}},
\end{align}
where $E_{0z}$ and $H_{0z}$ are the initial values of the electric and magnetic fields at the boundaries, respectively, and $\vec{k}_0$ is the propagation constant. A set of monochromatic plane waves is excited for each frequency in the simulation domain. This way, a point defect is introduced in the topologies characterized in previous sections. The resulting transmittance results are depicted in Fig.~\ref{transmittancesqtri}. 

When the size of the cluster is enlarged, by adding subsequent periods to the lattice, the transmittance spectra clearly reveals a sharp peak centered at the defect mode frequency. When one considers a larger cluster, the defect mode gets narrower, i.e. it converges to the previously discussed periodic case (Fig. ~\ref{transmittancesqtri}). Additionally, the defect mode frequency is shifted due to the overall dielectric percentage change in the finite cluster which is indeed very convenient since it allows one to tune the resonance frequency according to dielectric index changes. 

\begin{figure*}
	\includegraphics{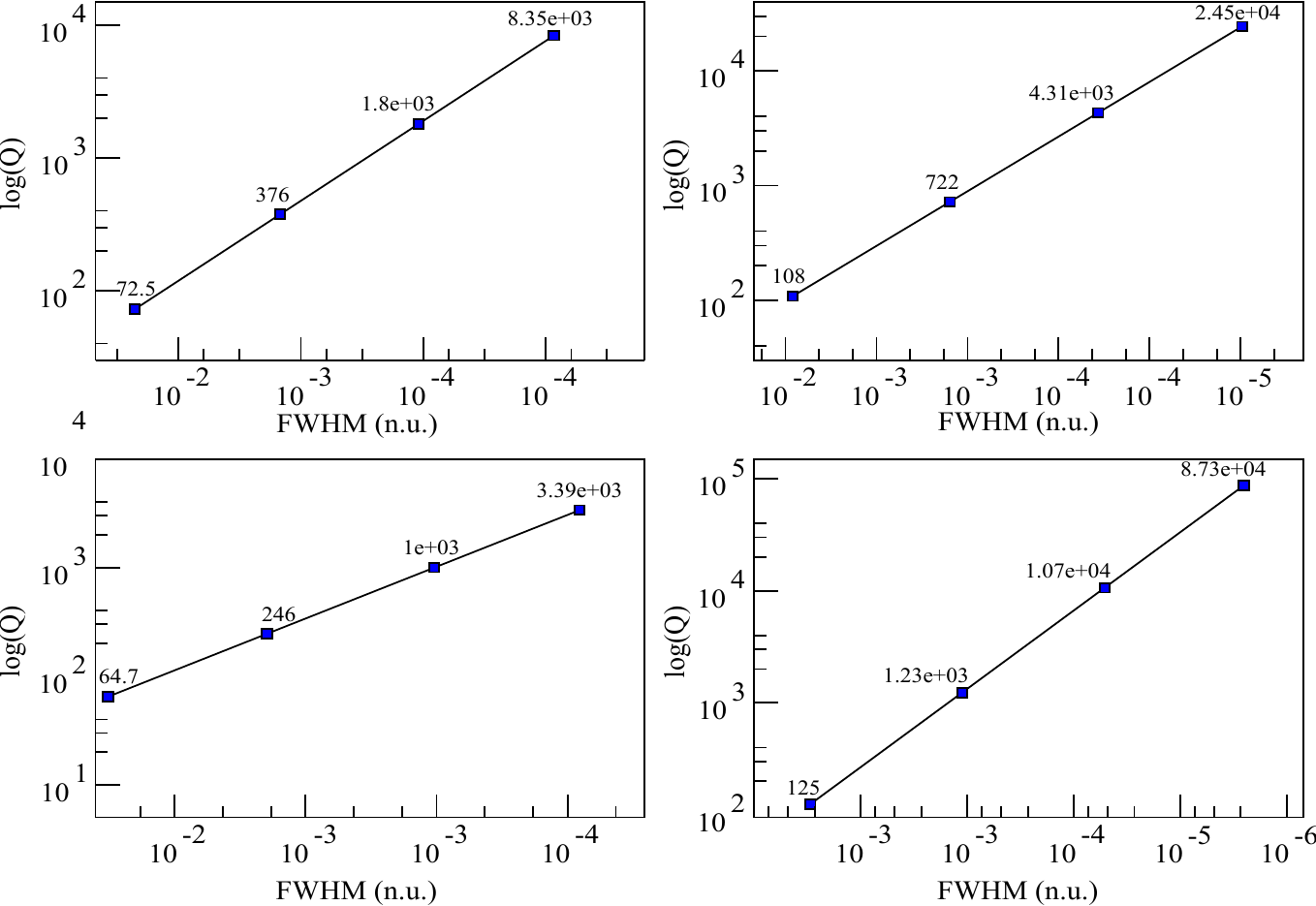}
	\caption{\label{Q}(color online) Top: quality factor estimation for a number of different sized square clusters of $0.38 \textit{a}$ radii dielectric rods (left) and $0.17 \textit{a}$ rods (right). Down: analogous computation results obtained for triangular arrangements.}
\end{figure*}

For the particular case of a square arrangement of dielectric rods with radius $0.38\textit{a}$ and dielectric constant of $9$, wherein its dimensions have been gradually modified, the energy stored in the defect mode centered around $\frac{\textit{a}}{\lambda}=0.4686$ is exchanged with the upper band energy situated around $\frac{\textit{a}}{\lambda}=0.49$. 
The quality factor has been calculated by means of the transmittance response for each crystal, since by definition
\begin{equation}\label{eq:qualityfreq}
  Q = \frac{f_{c}}{FWHM},
\end{equation}
where $f_{c}$ represents the central frequency of the defect mode and FWHM the  difference between the two values in the transmittance function at which the transmittance is equal to the half of defect mode maximum amplitude. For each quality factor calculation, the FWHM has been accurately determined by means of the Brent's algorithm \cite{Brent}, which combines bisection, regula falsi, and inverse quadratic interpolation methods for root finding. In contrast with the accepted wisdom, this method has proved to be highly reliable and more efficient than using FDTD, as we explain below in more detail. As the amount of dielectric rods surrounding the defect increases, the rate of the energy loss within the cavity relative to the energy confined in it decays exponentially. Figure~\ref{Q} depicts the quality factor increment for these square and triangular clusters. According to that figure, the defect mode bandwidth decreases exponentially with the addition of new periods due to the strengthening of the band gap effect, which is the main contribution to the enhancement of the quality factor. The  onset of the leakage mechanism transforms the localized modes transmittance spectra into distorted Lorentzian peaks. These peaks progressively tend to a Lorentzian curve when the defect mode gets closer to the mid-gap frequency and its FWHM gets narrower whenever $N$ increases. 

\section{Time domain approach: FDTD}\label{FDTD} 

As FDTD is the most popular method for the theoretical description of light propagation in these systems, we  used FDTD in order to assess and compare the \textit{Q} factor calculations attained so far using FEM. For the present manuscript, the FDTD simulations were performed using MEEP \cite{FDTD}, a freely available implementation package of this method. The success of FDTD method is due to its flexibility and to its adaptability to irregular or aperiodic geommetries. More generally, the FDTD method can compute a large number of frequencies at once and even extract modes of the spectrum. 

In FDTD space and time is divided into a finite rectangular grid and then fields are evolved in time using discrete steps. However, FDTD has serious limitations when the computational domain is finite whilst FEM convergence time is rather insensitive to this fact. In FDTD the computational domain must be terminated with some boundary conditions as it is the case in the FEM and, in order to simulate open boundaries in finite clusters, a wave absorbing mechanism, such as the split-field PML proposed by Berenger \cite{Berenger} has been used. Such an artificial region is needed so as to absorb outgoing spurious waves from the computational grid rather than reflecting them back into the photonic crystal. Moreover, quality factor calculations are very sensitive to the size of the computational grid and, thus, if the injected pulse experiences spurious reflections from the domain boundaries, radiated normal-incident waves will not be the dominant ones and they will be mixed with reflected waves. MEEP implements Uniaxial Perfectly Matched Layers (UPML) \cite{UPML,PML_problem} in order to absorb outgoing spurious waves from the computational. Along this artificial medium, the PML is expressed in terms of effective anisotropic $\epsilon$ and $\nu$ \cite{MEEP}. 

\begin{figure*}
	
	\includegraphics{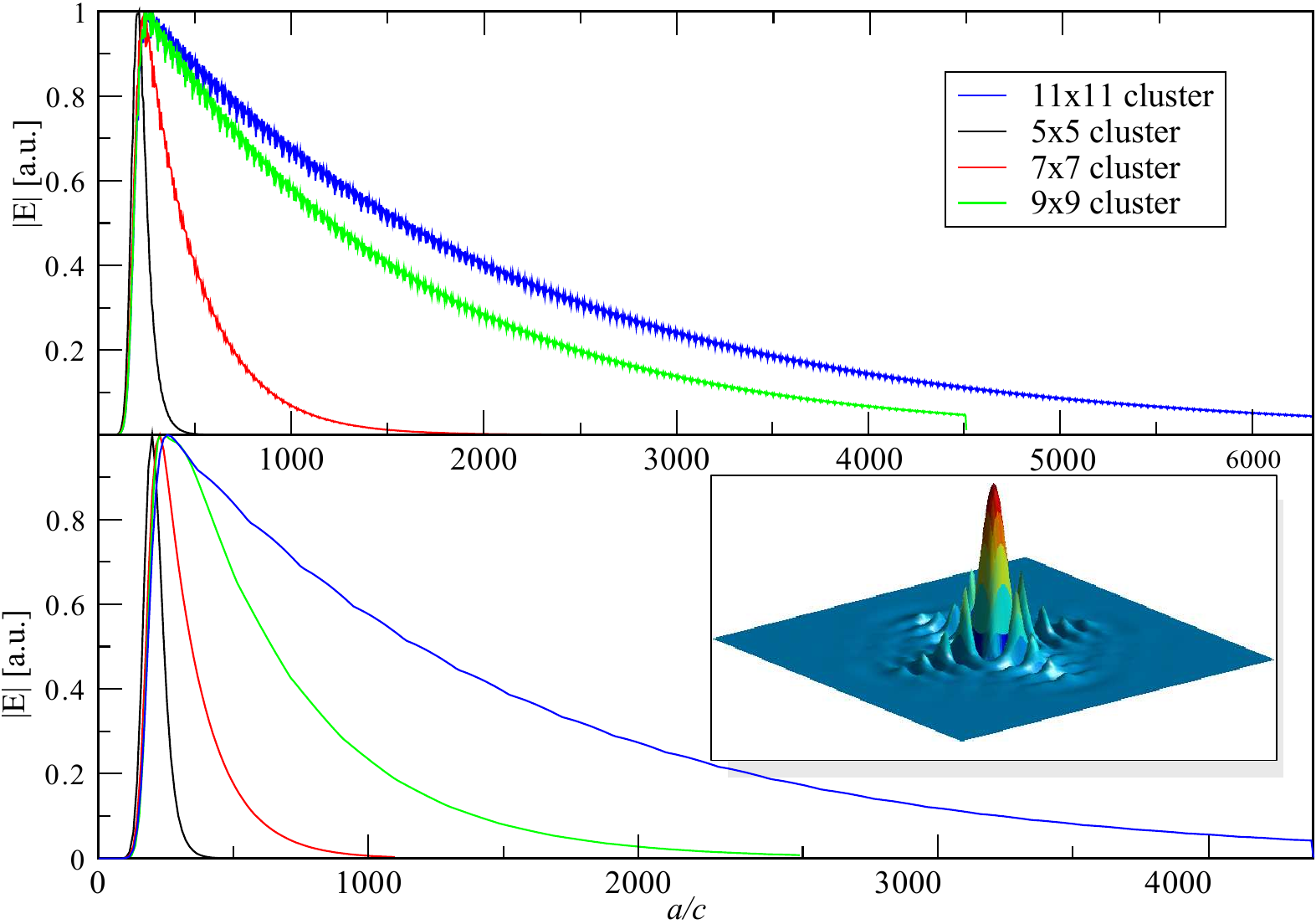}%timeresultssq
	\caption{\label{timeresultssq} (color online) Top: evolution of the electromagnetic field inside the defect rod for a $0.38 \textit{a}$ arrangement and for different cluster dimensions. A point source is excited in the cavity but after some periods the source is extinguished and although the electromagnetic field still remains in the cavity it experiences an exponential decay. Down: analogous computation results obtained for triangular defect cluster.}
\end{figure*}

For the present FDTD computation, a point Gaussian source has been placed inside the cavity. This excitation must be short enough (broad bandwidth) to excite the defect mode for each cluster. When this Gaussian source is switched on, the field grows and after some time this source is extinguished. Subsequently, a resonance effect occurs and the electromagnetic fields bounce back and forth for a limited amount of optical periods. Meanwhile, the energy trapped around the defect exhibits an exponential time decay (see Fig. ~\ref{timeresultssq}). At this point, if one takes into account the slowly varying component only, i.e., the envelope of the electric field norm, the quality factor is determined by the ratio of the stored power divided by the loss power after one cycle:

\begin{equation}\label{eq:qualitytime}
  Q =2 \pi \frac{\abs{E_{t}}^{2}}{\abs{E_{t}}^{2}-\abs{E_{t+T}}^{2}}.
\end{equation}

In this type of calculations, the proximity of the UPML to the cluster significantly determines the decaying factor of the existing resonance. In the FDTD calculations reported in Table \ref{result_table} an UPML of thickness $2\textit{a}$ surrounds each structure. Besides, the distance between the UPML layer and the edge of the computational grid must be adequately tuned in order to avoid an unphysical field decay. According to this, a distance of $0.8 \le d \le 1.1$ has been used. 
Sampling the electric field response with a period relative to the resonance frequency, results in a quasi-periodic step function that induces some uncertainties in the \textit{Q} factor determination if (\ref{eq:qualitytime}) is directly applied. These fluctuations could be due to the abruptly broken translational symmetry of the clusters. However, the marked decaying behavior of the electric field data can be easily filtered and interpolated. Therefore, by iteratively applying (\ref{eq:qualitytime}) to the interpolated sample field, an average \textit{Q} factor and the corresponding theoretical error is obtained.
Finally, in order to compare with previous \textit{Q} factor calculations obtained via FEM transmittance results and FDTD decaying field value relations, an harmonic inversion of time signals (\texttt{Harminv}) \cite{harminv} implementation has been used. This software package uses a filter diagonalization method (FDM) to find the deconvolution of sinusoidal functions near a given frequency interval. The simulation setup is the same as the one used in the standard FDTD calculations: a broad Gaussian point source located in the defect excites TE modes in the cavity and PML layers surrounding the cluster. In particular, a $\tfrac{a}{\lambda}=0.02$ pulse width Gaussian source centered at the resonant frequency of each structure has been excited. When the source vanishes, \texttt{Harminv} performs the signal processing of the fields in the cavity. This way, it identifies the frequencies and decay rates of the excited resonant modes. This method is a fast and accurate way to determine the \textit{Q} factors but still requires waiting until the fields have evolved and decayed to a certain small value. Computation time scales also with the cluster size, but this fact can often be overcome if symmetries are used to reduce the size of the computational cell. Besides, choosing a broad Gaussian source can yield to the appearance of spurious solutions but, at the same time, the source must be broad enough to excite the resonant frequency. 

It is important to stress the remarkable agreement between FEM results, FDTD and Harminv calculations of the \textit{Q} factor. Table \ref{result_table} summarizes the results obtained using the three methods described in this report. Two dimensional FEM calculation of \textit{Q} factors offers a very satisfactory agreement with FDTD  and Harminv results whilst providing substantial gain in solution robustness and efficiency. Indeed, FDTD is a powerful tool to calculate resonant frequencies and quality factors for complex cavity structures. However, it is very inefficient because one must discard many simulation cycles before reaching the stationary regime. Also, there are many heuristic factors that enter into the FDTD simulation process, such as the thickness of the UPML, the excitation source location and both its central frequency and width, that make the results from this method not as reliable as one would want. Moreover, when computing resonant modes in time domain especial care must be taken to avoid choosing an excitation source nearly orthogonal to the resonant mode because FDTD is likely to miss it, otherwise. In this regard, FEM can be seen as an efficient, reliable and more rigorous alternative to FDTD for the analysis of quality factors and resonant modes in complex dielectric material.

\begin{table*}[ht]
\begin {center}
%\caption{\label{result_table}Computation methods comparison square cluster $r=0.38\textit{a}$}
% title of Table

\begin{tabular}{ c r D{,}{\pm}{-1} r r D{,}{\pm}{-1} r }
\hline\hline

 &\multicolumn{3}{c}{Square lattice r=$0.38$\textit{a}}&\multicolumn{3}{c}{Square lattice r=$0.17$\textit{a}}\\

\hline
\multicolumn{1}{c}{Cluster size}
& \multicolumn{1}{c}{FEM} & \multicolumn{1}{c}{FDTD}& \multicolumn{1}{c}{Harminv}& \multicolumn{1}{c}{FEM}& \multicolumn{1}{c}{FDTD}& \multicolumn{1}{c}{Harminv}\\

%Cluster size&FEM&FDTD&Harminv&FEM&FDTD&Harminv\\
% \\&FEM&FDTD&Harminv&FEM&FDTD&Harminv\\
\hline

5& 72 & 80 , 1 & 88 & 108 & 113 , 13 & 107 \\
7& 376 & 350 , 3 & 347 & 722 & 727 , 6 & 723 \\
9& 1800 & 1773 , 24 & 1765 & 4310 & 4312 , 31 & 4308 \\
11& 8350 & 7677 , 83 & 7674 &24500 & 24561 , 74 & 24559 \\

\hline

 &\multicolumn{3}{c}{Triangular lattice r=$0.38$\textit{a}}&\multicolumn{3}{c}{Triangular lattice r=$0.17$\textit{a}}\\
\hline
5& 97 & 107 , 2 & 103 & 134 & 130 , 1 & 126 \\
7 & 358 & 320 , 1 & 316 & 1280 & 1171 , 8 & 1166 \\
9 & 1240 &  1211 , 5 & 1209 & 11100 &  11241 , 42 & 11238 \\
11 & 5270 & 5656 , 74 & 5650 & 90200 & 83903 , 84 & 83892 \\ 
\hline
\hline
\end{tabular}
\caption{\label{result_table}Quality factor calculations for diferent cluster sizes in square and triangular lattice rod-type PC single-defect cavities using FEM, FDTD, and harmonic inversion methods. }
\end{center}
\end{table*}

\section{Conclusions}\label{conclusions} 
The present manuscript reports a comprehensive study of the photonic properties of several two-dimensional photonic crystals and finite clusters by using the finite element method. The main result coming out from these calculations is that the FEM allows one to reproduce the results obtained with the well-known plane wave and FDTD methods, but it has many advantages not present in the others. In contrast with frequency-space based approaches, FEM also deals in a natural way with finite clusters and aperiodic structures of arbitrary complexity. To prove this, we have calculated the band structure of periodic photonic crystals based on the square and triangular lattices. It is found that the band structures calculated in this way are almost indistinguishable of those calculated with the well known MPB package. Also, the modes calculated with FEM closely resemble those calculated with the PWE method. It is noticeable that using a coarser discretization FEM results are more acurate than the ones given by PWE. 

Moreover, the transmission coefficients of a number of finite clusters of the aforementioned lattices were calculated along different directions in reciprocal space and for both TE and TM polarizations. The features (such as the position and width of the photonic gaps) of these transmittances agree quite well with the band structures and once again the accuracy of the band and mode calculations is very good when compared with the PWE method. Therefore, these results demonstrate that the FEM method can be a very useful general purpose method for investigating photonic crystals. This fact is further stressed by the results obtained for PCs containing a single defect, where DOS and dispersion diagrams have been calculated. There, experimental results obtained by McCall and coworkers, MPB calculations, and FEM results fully agree. As seen in these cavities, the confinement of the electric field amplitude strongly depends on the band gap strength of the underlying structure and, thus, wider gap geometries support higher mode localization. In addition, former calculations have been reproduced for finite length point defect clusters. In this context, the localization of light around the defect region has been quantified by accurately determining the quality factor using FEM, FDTD, and \texttt{Harminv} procedures for different cluster dimensions.

FEM is proven to be an effective and stable tool for point defect cluster quality factor calculation, wherein for each quality factor calculation, the FWHM has been accurately determined by means of Brent's algorithm. This technique permits to determine the essential information needed for \textit{Q} factor calculation in a speedy and computationally effective way. In addition, the leakage mechanism of PC cavities transforms the transmittance spectra into an almost Lorentzian peak and, therefore, with few transmittance calculations one can obtain the entire point defect transmittance response. It is noteworthy, with regards to the the point defect PCs addressed above, the fact that the three methods accurately reproduce the \textit{Q} factor for each topology. However, FDTD based methods have serious drawbacks. On the one hand, the width and the proximity of the absorbing layers significantly determine the decaying behavior of the trapped fields inside the cavity, and so, the parameters of the PMLs must be carefully chosen for each structure. In the aforementioned point defect structures, the source must be placed inside the cavity, which in some cases can be an unrealistic situation. Furthermore, the bandwidth of the source must be set intuitively so as to excite only the defect mode. On the other hand, we believe that among these methods FEM is desirable for the calculation of \textit{Q} factor due to its high numerical efficiency and stability because in FDTD, after the source is extinguished, one must wait an uncertain amount of time until the fields evolve and decay. In fact, FDTD gives quite accurate values for both the resonant frequency and the \textit{Q} factor but, for higher \textit{Q} values, the slope of the electromagnetic field inside the defect is nearly zero and hence, convergence time and and numerical errors increase drastically.

To summarize, the results presented in this manuscript demonstrate that the finite element method is an effective, stable, robust, and rigorous tool to study light propagation and confinement in both periodic and aperiodic dielectric photonic crystals. Furthermore, we expect that these advantages can be extrapolated to systems in which the optical constants are frequency dependent, such as hybrid metallo-dielectric photonic crystals.

\begin{acknowledgments}
We would like to thank the Basque Government for financial support under the SAIOTEK 2012 programme (ref. SIGMA) and the Grupos Consolidados 2006--2012 programme (Grant No. IT-331-07).
\end{acknowledgments}

%%%%%%%%%%%%%%%%%%%%%%% References %%%%%%%%%%%%%%%%%%%%%%%%%

\end{document}